\documentclass[twocolumn,showpacs,preprintnumbers,amsmath,amssymb]{revtex4}
\usepackage{graphicx}

\begin{document}

\title{Clogging of soft particles in 2D hoppers}

\author{Xia Hong}
\author{Meghan Kohne}
\author{Mia Morrell}
\author{Haoran Wang}
\author{Eric R. Weeks}
\affiliation{Department of Physics, Emory University,
Atlanta, GA 30322, USA}

\date{\today}

\begin{abstract}
Using experiments and simulations, we study the flow of soft
particles through quasi-two-dimensional hoppers.  The first
experiment uses oil-in-water emulsion droplets in a thin sample
chamber.  Due to surfactants coating the droplets, they easily slide
past each other, approximating soft frictionless disks.  For these
droplets, clogging at the hopper exit requires a narrow hopper
opening only slightly larger than the droplet diameter.  The second
experiments use soft hydrogel particles in a thin sample chamber,
where we vary gravity by changing the tilt angle of the chamber.
For reduced gravity, clogging becomes easier, and can occur for
larger hopper openings.  Our simulations mimic the emulsion
experiments and demonstrate that
softness is a key factor controlling clogging: with stiffer
particles or a weaker gravitational force, clogging is easier.
The fractional amount a single particle is deformed under its own
weight is a useful parameter measuring particle softness.  Data
from the simulation and hydrogel experiments collapse when compared
using this parameter.  Our results suggest that prior studies
using hard particles were in a limit where the role of softness
is negligible which causes clogging to occur with significantly larger
openings.
\end{abstract}


\maketitle

\section{Introduction}
\label{intro}

Flowing sand differs qualitatively from flowing fluid and
understanding the differences leads to interesting physics
\cite{jaeger96,forterre08}.  A dramatic difference is seen in the
gravity-driven flow of sand out of a hopper:  when the exit opening
from a hopper is small, the sand can clog at the hopper exit
\cite{zuriguel14,zuriguel14pip}.  The
existence of a critical exit opening size of 3-6 particle diameters
has been long known
\cite{deming29,brown58,beverloo61,nedderman82,sheldon10,aguirre10,wilson14}.
Even when the hopper opening is slightly larger, and clogs do
not form, the flow is influenced by the possibility of clogging:
for example, there are fluctuations of the flow rate of the sand
\cite{fowler59,hong92,vivanco12,zuriguel14}.  The mean flow rate is a function
of the difference of the opening size to the critical size for
clogging, a result often attributed to Beverloo \cite{beverloo61}
although mentioned by earlier authors as well \cite{deming29};
the history is discussed in Ref.~\cite{nedderman82}.  In this
sense, understanding what happens when hoppers clog -- and
the size of the opening that causes clogging -- is crucial for
understanding the flow properties when the opening is 
larger than the critical size
\cite{beverloo61,nedderman82,zuriguel14pip}.  We note that some experiments
suggest that clogging does not have a critical size but rather
becomes exponentially unlikely as the hopper opening increases
\cite{to05,janda08,thomas15}; nonetheless, it's clear that 
understanding the flow properties requires understanding the
clogging probability.

The clogging process itself is due to arch formation
at the hopper exit \cite{hong92,to01,janda08,to02}.  The difficulty of forming
large arches is the reason why hoppers do not clog when their
exit opening is sufficiently large \cite{to01}.  
Friction may be important for the formation of
these arches \cite{to01}, and more generally it has long
been seen that friction influences hopper flow to an extent
\cite{deming29,franklin55,fowler59,nedderman82,vivanco12}.
However, it was unclear exactly how friction played a role --
friction influences the angle of repose \cite{franklin55} and the
packing density \cite{beverloo61}, for example, but it was unclear
which of these (if either) influences the flow rate or clogging.
Another experiment studied
the shapes of arches formed in 2D granular hoppers, finding
that these shapes differed somewhat from simulated frictionless
arches \cite{garcimartin10}; static friction allowed some arches
to form that would be unstable in a frictionless situation.

The role of particle softness has been less studied.  One experiment
studied the flow of foams, and showed that the softness of the
bubbles influenced the flow \cite{bertho06}.  In this case, there
was no static friction.  Due to the ability of bubbles to deform,
clogging required the exit orifice to be smaller than the mean
bubble size, and this profoundly changed the flow rate at larger
exit orifice sizes \cite{bertho06} as compared to the granular
Beverloo flow law \cite{beverloo61}.  One recent experiment used
repulsive magnetic particles in a quasi-2D hopper and reported
clogging for small orifices, but did not systematically study
clogging \cite{lumay15}.  In that work, the particles repelled
each other at moderate separations, and so it was not clear how
the clogging related to the particle size (or even how to define
that size).  A pair of papers simulated the flow of softer
frictional particles through 2D hoppers, and found that as the driving
force increased by a factor of $10^4$ there was a mild decrease in
clogging \cite{arevalo14,arevalo16}.  A key result was that as
the driving force was decreased toward zero, there was a clear
finite probability for clogging, suggesting that geometric
effects are important \cite{arevalo16}.

\begin{figure}
\includegraphics[width=8cm]{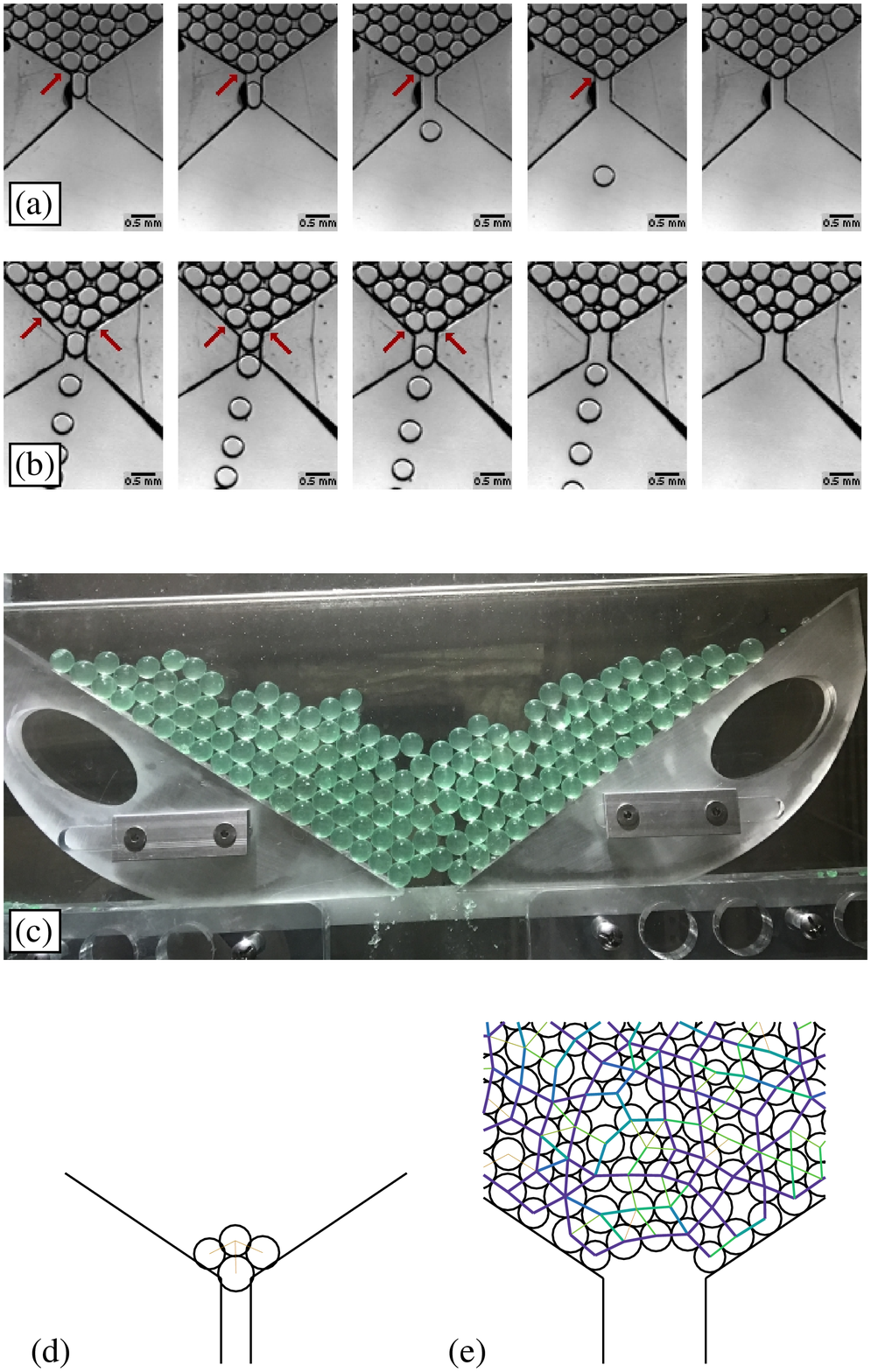}
\caption{
(a) Clogging of oil droplets in water passing through a hopper
with exit width $w/d=0.81$ with mean droplet diameter
$d=370$~$\mu$m.  
The images are each 10~s apart, except for the final image which is
50~s later.
(b) Clogging with an arch composed of two oil droplets
with $w/d=1.00$ where $d=410$~$\mu$m.  
The images are each 5~s apart,
except for the final image which is 30~s later.
In (a) and (b) the arrows indicate
the droplet(s) that will clog the opening.  
(c) Photograph of the hydrogel experiment in a clogged state.  The
sample chamber is tilted at an angle $\theta = 10^\circ$ from the
horizontal and the opening width is $w=28.8$~mm $=2.2d$ in terms
of the mean particle diameter $d=1.31$~mm.
(d,e) Simulated clogging arches.  The parameter values are 
(d) $g/F_0=10^{-1}$, $w/d=1.74$, 4 droplets left in hopper; 
(e) $g/F_0=10^{-4}$, $w/d=6.0$, 708 droplets left in hopper.  
The colored lines indicate contact forces between the droplets,
relative to the gravitational force acting on an isolated droplet
of the mean size.  The thickest (purple) lines correspond to
forces 8 or more times larger than the reference force.  
}
\label{pictures}
\end{figure}

In this paper we study clogging in flow out of a hopper using 
two quasi-two-dimensional
experiments with soft nearly frictionless granular materials,
and also simulations of soft frictionless particles.  Our first
experiment is an emulsion composed of oil droplets in water,
stabilized by a surfactant, as shown in Fig.~\ref{pictures}(a,b).
The droplets are sandwiched between two parallel pieces
of glass so that they are deformed into pancake-like disks
\cite{desmond13}.  In our experiments droplets only
clog when the hopper opening is less than two diameters wide.
Clogging arches involve only one or two droplets.  Our second
experiment uses soft hydrogel particles in a thin sample chamber
as shown in Fig.~\ref{pictures}(c).  The influence of gravity is
varied by changing the tilt angle of the chamber.  Reducing the
influence of gravity enhances clogging, allowing clogs to occur
for hopper openings ranging from 1.5 - 2.5 diameters wide.
Our results are a strong contrast to prior experimental
results which used hard frictional granular particles, which
saw larger arches and which clogged at larger opening sizes
\cite{brown58,tang09,to01,janda08,vivanco12}.  To vary the particle
softness, we conduct simulations using the Durian bubble model
\cite{durian95} with the particles a few orders of magnitude softer
than previously studied \cite{arevalo14,arevalo16}.  The simulation
results show that the softness of the experimental particles
explains the difficulty of clogging; Fig.~\ref{pictures}(d) shows
a situation where the particles are quite soft, and
(e) where they are harder.  Our simulation results suggest that
making the particles harder (or reducing gravity) can potentially
recover the previous experimental results for hard particles.
This demonstrates the importance of softness to
the clogging process, and shows that flowing particulate materials
behave qualitatively different when the particles are easily
deformable by the flow.

\section{Methods}

\subsection{Emulsion experiment}

\begin{figure}
\includegraphics[width=8cm]{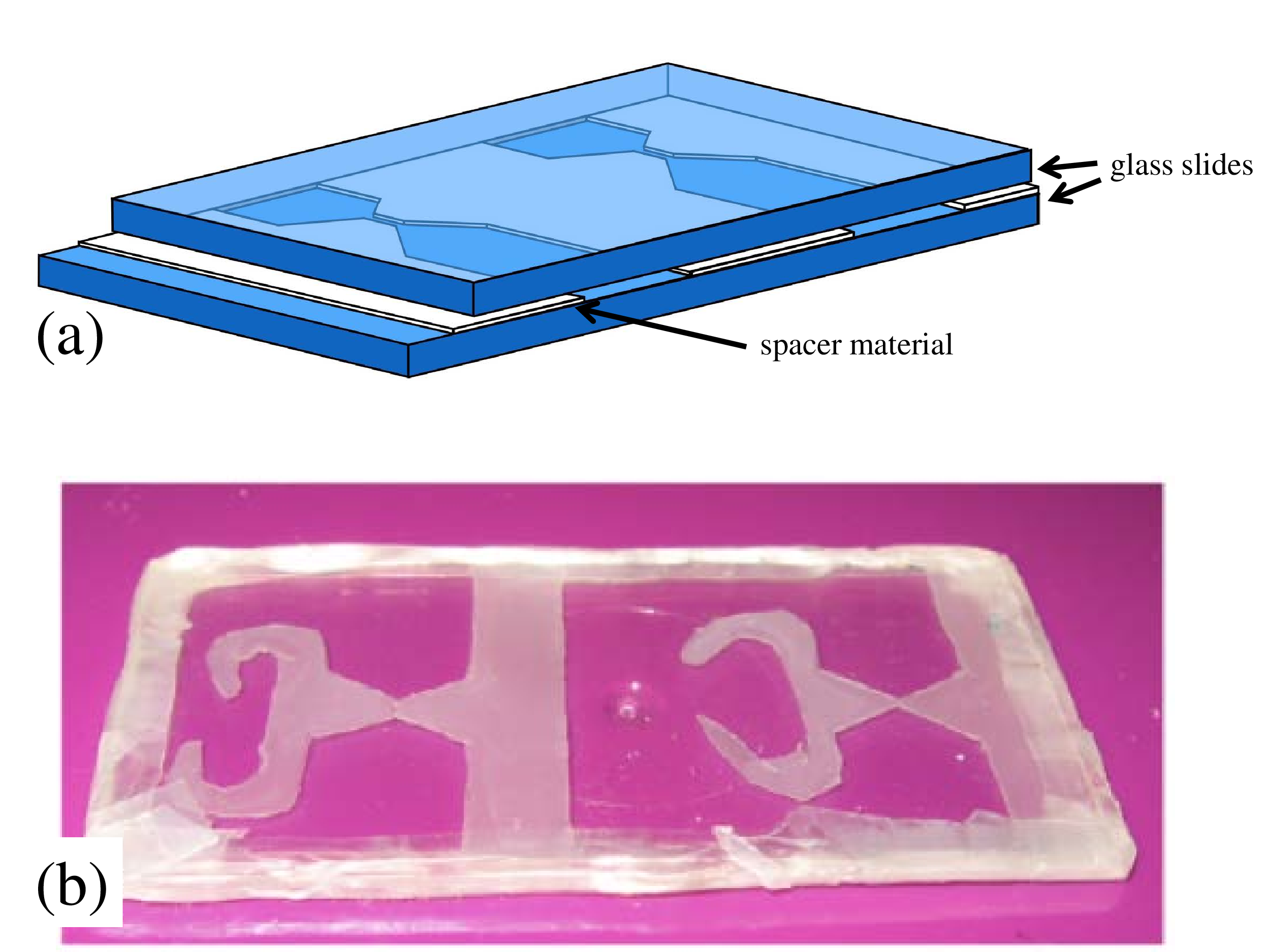}
\caption{
(a) Sketch of a sample chamber for our emulsion
experiments.
(b) Photograph of a typical sample chamber constructed from
parafilm.  This slide contains
two separate hopper chambers that are not interconnected.
}
\label{chamber}
\end{figure}

Our samples are oil-in-water emulsions prepared by a standard
co-flow microfluidic technique \cite{shah08}.  In this technique,
mineral oil (Fisher Scientific O121-1, density $\rho_{\rm
oil}=0.83$~g/mL) is injected into a flowing stream of distilled
water and surfactant.  We use Fairy dishwashing detergent at mass
fraction 0.025 as the surfactant, as has been done in previous
work \cite{desmond13}.  The microfluidic technique produces
droplets of a desired size with 3\% polydispersity.
We control the size of the droplets by varying the flow rates
of the oil and water in the microfluidic device.  Typically we
make droplets $\sim 200$~$\mu$m in diameter.  In some cases we
mix together two batches of droplets with different sizes, but
for most of our results we study samples composed of a single
batch of droplets.  Sometimes the emulsion gets sheared when we
add it to the sample chamber, resulting in a few droplets that
are unusually small, or the coalescence of droplets so that some
are unusually large.  Examples of each can be seen in
Fig.~\ref{bigsqueeze}.

Each sample chamber is a sandwich of a spacer between two glass
slides, as shown in Fig.~\ref{chamber}(a).  The spacer material is
either transparent plastic film ($\approx 120$~$\mu$m thickness)
or parafilm ($\approx 130$~$\mu$m thickness).  For each of these,
the spacer material is cut into a desired shape using scissors.
We briefly put the parafilm chambers onto a hot plate to slightly
melt the parafilm to seal the chamber.  In each case, after
the initial preparation, the sample chambers are additionally
sealed with epoxy to prevent leakage or evaporation.  As we use
scissors and position the spacer materials onto the slides
by hand, often the sample chambers are imperfect.  However,
given the simplicity and rapidity of making these chambers, we
simply select the best sample chambers to use in our experiments,
where the hopper exit is adequately shaped.  Examples are shown
in Figs.~\ref{pictures}(a,b) and \ref{bigsqueeze}.  The hopper
angles are set to be $32-35^\circ$, close to To {\it et al.}'s experiment
with an angle of $34^\circ$ \cite{to01}.

Should droplets flowing through the hopper clog, we need a way to
unclog the system and get all of the droplets back to the entrance
side of the hopper.  We design our sample chambers with a side
channel as shown in Fig.~\ref{chamber}(b).  This allows the sample
chamber to be tilted and gives a path to move droplets from one
side of the hopper to the other.  The ``C" shape on the left side
of the individual chambers shown in Fig.~\ref{chamber}(b) is to
collect and hold any air bubble that might be present after the
emulsion is added to the chamber.

Given the fairly large size of the droplets, we use a CCD camera
and a macro-zoom lens to view our experiments, back-lighting
the sample chamber.  Jammed hoppers can also be seen by eye,
which makes it possible to collect statistics without the camera.
Video microscopy is used to count the number of droplets within a
sample chamber, and to get an accurate measurement of the hopper
angle of each chamber.

\subsection{Physics of flowing emulsions}

While we are motivated by experiments on granular hopper flows,
as described in Sec.~\ref{intro}, there are several differences
in our emulsion experiment.  These differences are described in this
section.

A superficial difference is that the density of the mineral oil
droplets is smaller than water, so our droplets float upward due
to gravity.  To make easier conceptual comparison with granular
hoppers, we rotate all of our photographs so that the droplets
are moving downward, for example Fig.~\ref{pictures}(a,b).

\begin{figure}
\includegraphics[width=8cm]{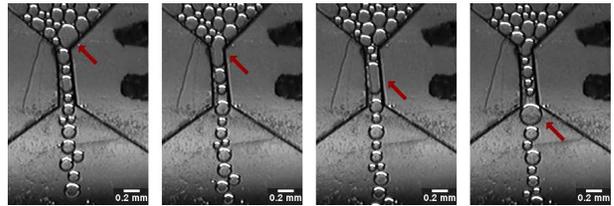}
\caption{This image sequence shows how a big droplet (marked with
the arrow)
can deform and squeeze through the hopper exit, if the surface
tension is too low.  The images are shown at 10~s time intervals.
}
\label{bigsqueeze}
\end{figure}

A second difference is that our droplets are soft and
deformable.  The original work by To {\it et al.} used steel
disks \cite{to01,to02,to02b,to05}, some authors use solid
spheres \cite{janda08,zuriguel14,thomas15,garcimartin10},
and others use slightly deformable photoelastic disks
\cite{tang09,vivanco12}.  Our droplets are significantly
more deformable.  In the absence of external forces, a droplet would
be spherical due to surface tension.  However, in our experiment,
droplets could potentially decrease their gravitational potential
energy by deforming to squeeze through the hopper.  This is
indeed what happens if the surface tension is too small, or if the
droplets are too large: an example is seen in Fig.~\ref{bigsqueeze}.
As the gravitational energy over a length scale $d$ scales with
droplet diameter as $d^4$ while surface energy scales as $d^2$,
larger droplets will prefer to deform to reduce their gravitational
energy \cite{weisskopf86}.  Accordingly, to study clogging in
our hoppers, we use a low amount of surfactant to keep
the surface tension high, and also we use smaller droplets.
This prevents the problem seen in Fig.~\ref{bigsqueeze}.  Were we
to use large droplets, they would still clog if the hopper opening
was sufficiently narrow, but this would then be entirely a surface
tension effect rather than a study of clogging.

A third difference between our experiments and the prior granular
experiments is that our oil droplets move through a viscous
background fluid (water, viscosity $\eta \approx 1$~mPa$\cdot$s).
The mineral oil droplets are themselves viscous ($\eta \approx
20$~mPa$\cdot$s) and experience viscous drag with the glass slides.
The droplets contact the glass slides with a contact angle of
$19^\circ$ \cite{desmond13}; in other words, there is no lubricating
water layer between the droplets and glass.  The viscous drag
on the droplets means that they move slowly with a free fall
velocity of $U = 0.20-0.25$
droplet diameters per s depending on the conditions.  This is
in contrast to the granular experiments where particles spill out
of the hopper quite rapidly \cite{tang09}.  This in principal might
make clogging easier, as droplets moving toward the hopper exit
have less inertia.  The Reynolds number ${\rm Re} = \rho \eta_{\rm
water} U / d$ is about 0.25 indicating
inertia is relatively unimportant even for the fastest moving
droplets.  Of course, prior 2D granular experiments
have some viscous drag from air, and also experience some sliding
friction against their confining walls \cite{to01,janda08}.


A fourth difference is that in a granular container, the pressure
is independent of depth (apart from near the free surface at
the top, and at the bottom near the exit).  This is known as the
Janssen effect \cite{janssen1895}, and is due to the frictional
forces acting on the particles from the container sidewalls
\cite{shaxby23,nedderman82}.  Due to our droplets not
having static friction, we would not expect the Janssen effect
to be present in our experiment.  
The lack
of a Janssen effect was confirmed by an earlier experiment by
our group, which found the internal pressure within a similar
quasi-2D emulsion pile depended on depth in a tall container
\cite{desmond13}.  This also is similar to a granular hopper
experiment using submerged particles \cite{wilson14} which did not
find a Janssen effect.  Accordingly, we might expect that clogging
is less likely at the start of our experiment when the weight of
the pile can more easily break an arch.  On the other hand, the
density mismatch between the oil and water is only $\Delta \rho =
0.17$~g/mL, so the gravitational forces acting on our droplets
are small albeit necessary for driving the hopper flow.  

To be clear, the Janssen effect is thought to be irrelevant
for understanding hopper flow.  For example, one experiment
removed the influence of gravity and provided strong evidence the
Janssen effect is unrelated to clogging and flow rates through
hoppers \cite{aguirre10}.  It is well known
that granular hopper flow is independent of the amount
of material in the hopper, as long as the amount of material is
above some minimal threshold \cite{beverloo61,aguirre10,thomas15}.
In contrast, that should not be the case in our experiments (and
this contrast is confirmed by our simulations).
As the weight above the droplets at the exit
decreases, the probability of clogging increases.  In other
words, our experiment cannot be treated as in steady state,
in contrast to granular hoppers
\cite{thomas15,zuriguel14pip}.
For granular experiments, the existence of a steady state
allows one to focus on the amount
of material flowing out between clogs, using some method of unjamming a clog
\cite{thomas15,garcimartin10}.  In contrast, our experimental
protocol is based on To {\it et al.} where we study the probability
for the hopper to completely drain for a fixed initial number of
droplets \cite{to01,to02}.

\subsection{Hydrogel particle methods}
\label{hydrogel}


We use soft hydrogel particles for a second series of clogging
experiments.  The particles are a polyacrylamide gel (green
water beads, purchased from Gift Square D{\'e}cor, Amazon.com).
As purchased they are dry spheres around 1~mm diameter.  We swell
these in distilled water for 24 hours.  They are fairly
polydisperse, so we sieve the swollen particles.  After sieving,
the mean particle diameter is 13.1~mm, with a standard deviation
of 0.5~mm.  We place these particles in an acrylic hopper chamber
with thickness 17.0~mm so that the particles are constrained to a
quasi-2D geometry, as shown in Fig.~\ref{pictures}(c).  The particles
start in an upper storage chamber which has a bottom metal plate
inserted holding the particles in that chamber.
We initiate the experiment by rapidly removing that plate by
hand, allowing the particles to fall downward toward the hopper.
An identical storage chamber is placed below the hopper to contain
all particles that fall through the hopper.

The sides of the hopper are at $34^\circ$ angles to match the
emulsion experiments.  The opening width is adjustable; prior to
each experiment, the hopper blocks are pushed together against an
inserted plastic block of the desired width.  If the experiment
clogs, we move the hopper walls apart to drain the system and
then reset the walls to the correct opening width.

The entire system is mounted on a horizontal axle so that we can
rotate the apparatus to any angle $\theta$ relative to the
horizontal.
This allows us to vary the component of gravity in the plane of the
hopper by a factor of 6, from full gravity ($\theta = 90^\circ$) to
reduced gravity ($\theta = 10^\circ$, thus $g = g_0 \sin 10^\circ
\approx 0.17 g_0$).  For $\theta < 10^\circ$ the particles can form a tall
pile in the bottom storage container which interferes with those
flowing out of the hopper exit.

We use a TA Instruments AR2000 rheometer with a parallel-plate
geometry to measure several physical properties of our hydrogel
particles.  We first measure the Poisson ratio.  This is done by
hand-cutting individual hydrogel particles into roughly cubical
shapes.  We then slowly compress the cubes with the rheometer using a
flat plate, and image the cubes from the side during the compression and
subsequent decompression.  The relation between vertical strain and
horizontal strain is linear, leading to a Poisson ratio of $\nu =
0.27 \pm 0.03$ (the uncertainty is the standard deviation of four
measurements).  This measurement is in agreement with a theory
predicting $\nu \approx 0.3$ for a polymer gel with the
Flory-Huggins $\chi \approx 0.5$ \cite{bouklas12} It is also not
far from the range of Poisson ratio values measured by a prior
experiment ($\nu = 0.38 - 0.49$) \cite{chippada11}.


We next find the Young's modulus of the hydrogel particles by compressing
individual spherical particles with the rheometer, which measures
the normal force as a function of the rheometer plate position.
The resulting relation between displacement and
compression force is well fit by the Hertzian force law.  From the
Hertzian fit and using the mean value for the Poisson ratio, the
Young's modulus is $E = 140 \pm 30$~kPa (the uncertainty is the
standard deviation of five measurements).  

To measure the
friction coefficient we attach acrylic disks to the rheometer
tool and base plate and
compress a pair of hydrogels placed symmetrically a distance $R$
from the rheometer axis.  The particles are each trapped in small
wells made from glue to prevent rolling.  The rheometer 
measures the torque $\tau$ required to
rotate the top acrylic disk with a given normal force $N$.  We compute
the friction coefficient due to the pair of particles from $\mu =
\tau / 2 N R$, finding $\mu = 0.006 \pm 0.004$, confirming that
the hydrogel particles are nearly frictionless.  This is the same
order of magnitude as prior measurements \cite{gong06}.
The variability is likely due both to heterogeneities of the
particles and also the variability of the contact, which can
sometimes trap water \cite{yamamoto14}.  Likely $\mu$ varies
within our clogging experiment; the
main point is that it is always small \cite{gong06}.  We did not
measure the hydrogel-hydrogel friction coefficient although prior
work found that it is no more than 0.03 \cite{dijksman13}.

Similar to our emulsion droplet experiments, the hydrogel particles
are softer than prior granular experiments, and so the pile of
particles has an internal pressure that acts like a hydrostatic
pressure:  the more particles in the hopper, the larger the force on
the particles at the exit of the hopper.  There is no
noticeable Janssen effect for the depth of filling we use, as
confirmed by a different experiment with similar hydrogel particles
\cite{ashour17}.  In contrast to our emulsion experiments, the
hydrogel particles fall through air, so there is less viscous
damping.

\subsection{Simulations}
\label{simulations}

Due to challenges in experimentally varying parameters such as
surface tension or gravity over a wide range, we also simulate
the hopper flow of emulsions.  This is done with the Durian
``bubble model'' \cite{durian95}, using the version presented in
Ref.~\cite{tewari99} that allows each particle to have a variable
number of nearest neighbors.  While the model was designed for
bubbles in flowing foams, it also works for emulsion droplets.
The simulation is strictly two-dimensional.  Each
droplet feels several forces.  First is a repulsive contact force acting
on droplet $i$ from each neighboring droplet $j$, modeled as
\begin{equation}
\label{repulsive}
\vec{F}_{ij}^{\rm contact} = 
F_0 \Big[\frac{1}{|\vec{r}_i - \vec{r}_j|} -
\frac{1}{|R_i + R_j|} \Big] \vec{r}_{ij},
\end{equation}
using the droplet radii $R_i$, their positions $\vec{r}_i$, and
the vector $\vec{r}_{ij} = \vec{r}_j - \vec{r}_i$.  The neighbors
$j$ are defined as those droplets for which $|\vec{r}_{ij}| <
R_i+R_j$.
$F_0$ acts like
a spring constant and conceptually is due to the surface tension.
In this model, rather than trying to deal with the droplet surface
energy directly via describing the deformed droplet surface,
droplets are treated as undeformed circles which repel each other
only when they overlap.  Neighboring droplets also exert viscous forces on each
other if they move at different velocities,
\begin{equation}
\label{viscous}
\vec{F}_{ij}^{\rm viscous} = b(\vec{v}_i - \vec{v}_j).
\end{equation}
To model our emulsion experiment, we add three additional forces.  First,
we add in a repulsive force from the hopper walls similar to
Eqn.~\ref{repulsive}, 
\begin{equation}
\vec{F}_{i}^{\rm wall} = 
F_0 \Big[\frac{1}{|\vec{r_i} - \vec{r}_{\rm wall}|} -
\frac{1}{R_i} \Big] \hat{r}_{i,{\rm wall}},
\label{walleqn}
\end{equation}
where $\vec{r}_{\rm wall}$ is placed at the closest 
point on a wall to the droplet, and
$\hat{r}_{i,{\rm wall}}$ is a unit vector pointing normal to the wall.
Similar to the droplet-droplet contact forces, $\vec{F}_i^{\rm
wall}$ only acts if a droplet overlaps with the wall, that is, if
it is within a distance $R_i$ to the wall.  Second, we add in a
gravitational force proportional to the mass of each droplet,
\begin{equation}
\vec{F}_{ij}^{\rm gravity} = -\rho g R^2_i \hat{y}
\label{graveqn}
\end{equation}
which points in the $-\hat{y}$ direction and introduces the 2D density
$\rho$ and acceleration due to gravity $g$.  Third, we add in a
viscous force between the droplets and the confining plates,
\begin{equation}
\vec{F}_{ij}^{\rm plates} = -c R^2_i \vec{v}_i,
\end{equation}
which enforces a terminal velocity (equal to $\rho g/c$) for freely
falling isolated droplets.  Finally, following the original bubble
model method \cite{durian95,tewari99}, we note that we are modelling
a regime where inertia plays no role, and therefore these forces
sum to zero for each droplet $i$:
\begin{equation}
\label{bubblemodel}
\sum_{j}[ \vec{F}^{\rm contact}_{ij} +
\vec{F}^{\rm viscous}_{ij}] +
\vec{F}^{\rm wall}_i +
\vec{F}^{\rm gravity}_i -
\vec{F}^{\rm plates}_i = 0.
\end{equation}
This can be rewritten as an equation for each droplet's velocity
$\vec{v}_i$ in terms of the positions and velocities of all the
droplets \cite{tewari99}.  

We simplify our simulations by setting $\rho=b=c=1$.  In practice,
the viscous forces in the simulations are typically quite small,
as the droplets flow slowly out of the hopper, and droplets
generally move in similar directions to their neighbors ($\vec{v}_i
\approx \vec{v}_j$).  We simulate 800 droplets with a Gaussian
radius distribution with mean $\langle R \rangle=1$ and standard
deviation $\sigma_R=0.1$.  We set $F_0=1$ for our simulations unless
otherwise noted, and vary $g$.  As viscous forces are so small, the
key control parameter is the nondimensional ratio $\rho g \langle R
\rangle^2/F_0$, which expresses the relative importance of gravity
to the contact forces between droplets.  Given $c=\langle R \rangle
= 1$ we will write this parameter as ratio $g/F_0$.

\begin{figure}
\includegraphics[width=8cm]{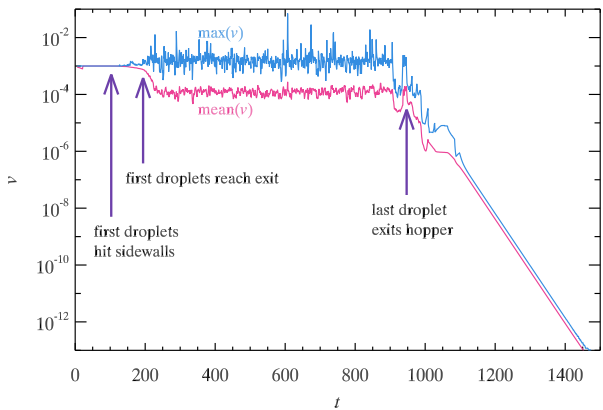}
\caption{
The maximum velocity (blue) and mean velocity (red) as a function
of time for a simulation run that clogs.  The maximum and mean
are taken over all droplets still in the hopper.
Until the droplets
first encounter the sidewalls of the hopper (first arrow), they
are in free fall.  Velocity fluctuations increase once the droplets
are flowing through the exit (second arrow), and the velocity then
decays to zero after the clog is formed (third arrow).  For this
simulation, the free fall velocity is given by $g=10^{-3}$ and
thus we end the simulation when the maximum velocity is below
$10^{-13}$.
This run corresponds to
a final state with a 4 droplet arch and 434 droplets left in the
hopper.
}
\label{velocityplot}
\end{figure}

Equation \ref{bubblemodel} is a first order differential equation;
to solve it we integrate using the standard fourth-order Runge-Kutta
algorithm.  There are several possible internal time scales in
our model:  $b\langle R \rangle / F_0, c\langle R \rangle^3/F_0,
b/\rho g \langle R \rangle,$ and $c\langle R \rangle / \rho g$.
Given our simplification $\langle R \rangle = \rho=b=c=1$ these
are two distinct time scales $1/F_0$ and $1/g$.  $1/F_0$ is the
time scale for two particles to push apart, limited by viscous
drag.  $1/g$ is the time scale for a particle to free fall a distance
$\langle R \rangle$ and in our simulations $(1/g) \gg (1/F_0) = 1$.
For the Runge-Kutta algorithm we use a time step of 0.1 for all
cases except for $g=10^{-4}$ (lowest
terminal velocity case) where the time step is 1.0.  We verified
that the results do not change for any cases when using a smaller
time step.  A clog is defined to have occurred in the simulation
when the maximum speed of all droplets is below $10^{-10}\rho g/c$.
A time course of the velocities seen in one simulation is shown
in Fig.~\ref{velocityplot}, showing that once the hopper clogs,
the velocities decay toward zero, and justifying our choice of
$10^{-10}g/c$ as a reasonable threshold for concluding that the
simulation has clogged.

The velocities change very slowly, so rather than solving
Eqn.~\ref{bubblemodel} for all velocities simultaneously, we use
the previous timestep's velocity values in Eqn.~\ref{viscous}
for the neighbor velocities.  Again, in practice, $\vec{F}_i^{\rm
viscous}$ is small compared to the other forces, so this is a
reasonable simplification.  For the simulations, the hopper angle
is $\theta=34^\circ$, and we use 800 droplets.

We initialize the simulations by placing droplets in random
positions above the hopper and with zero velocity.  Initially we
set the gravity in the opposite direction (away from the hopper
exit).  The droplets then move until they have reached positions
that minimize contact forces.  At that point, gravity is reversed
so that the droplets fall toward the hopper exit, much like the
way the emulsion experiments are conducted.  The resulting pile of droplets
above the hopper exit resembles the experimental conditions for both
emulsion and hydrogel experiments.  

\section{Results}
\label{results}

\subsection{Emulsion experiment}

\begin{figure}
\includegraphics[width=8cm]{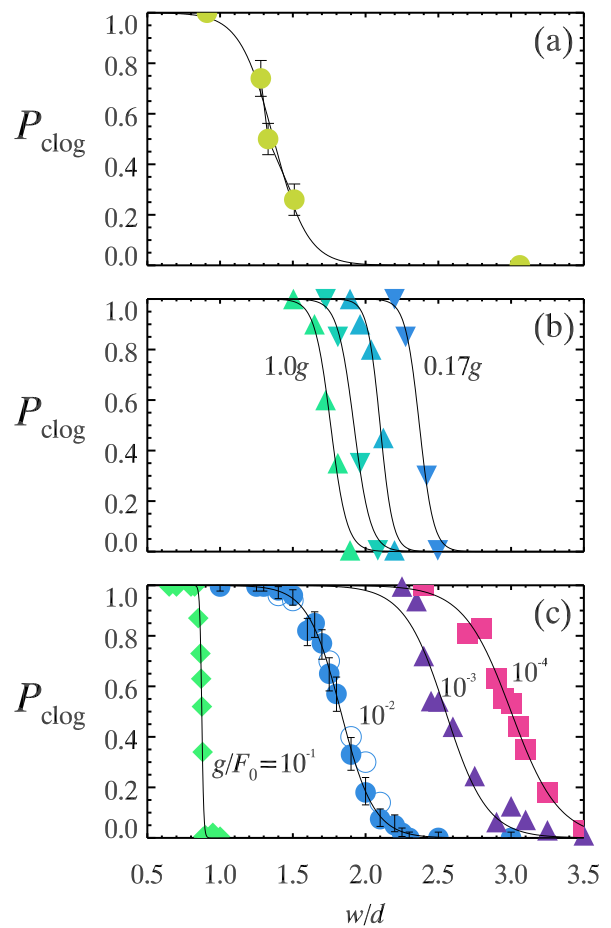}
\caption{The probability of clogging as a function of $w/d$, the
ratio of the hopper exit width $w$ to the droplet diameter $d$.
(a) $P_{\rm clog}$ for the emulsion experiments (data corresponding to
Table~\ref{pj}).  The solid line is a fit to the sigmoidal function
$P = [1 + \exp((w/d - a)/b)]^{-1}$ with $P = 1/2$ at $w/d=a=1.37$
and width $b=0.17$.  The error bars are the uncertainty due to
the finite number of trials ($n=50$) for a Poisson process.  
(b) Data from the hydrogel experiments with the influence of
gravity varied by setting the tilt angle at $\theta = 90^\circ,
43^\circ, 20^\circ, 10^\circ$ from left to right.  The lines are
sigmoidal fits with centers $a=1.76, 1.92, 2.10, 2.37$ and widths
$b=0.055, 0.057, 0.043, 0.048$ (from left to right).
(c) Simulation data, with $g/F_0$ decreasing from left to right as
labeled.  For these data, $F_0=1$, with the exception of the open
symbols for which $F_0 =10$ (and keeping $g/F_0=10^{-2}$ as
indicated).  Typical error bars are shown for
some of the data, based on the finite number of trials
($n=100$ for the simulations).  The lines are sigmoidal fits with
centers $a=0.87, 1.83, 2.55, 3.00$ and widths $b=0.0067, 0.11,
0.14, 0.16$ (from left to right).
}
\label{probplot}
\end{figure}

To determine clogging probabilities, we load our sample chamber
with 750 - 950 droplets.  We then let these droplets flow through
the sample chamber and observe if a clog forms.
We repeat this 50
times for each sample chamber to measure the clogging probability
$P_{\rm clog}$ (the fraction of experiments that clog).  
Figure \ref{probplot}(a) shows $P_{\rm clog}$ as
a function of the hopper exit width $w$ (normalized by the mean
droplet diameter $d$).  The most striking result is that the widths
at which clogging occurs are quite small.  At $w/d = 1.37$,
the droplets clog in half of the experiments, and for larger
openings,
clogging is never observed.  This is in stark contrast to the case
of hard frictional particles, which clog half of the time at $w/d
\approx 4$ \cite{to01}.

\begin{table}[t]
\begin{center}
\begin{tabular}{ccccc}
$w/d$ & $d$ & $N$ & $\theta$ & $P_{\rm clog}$ \\
\hline
0.30 & 237~$\mu$m & 867 & $32^\circ$ & 1.00 \\
0.91 & 202~$\mu$m & 947 & $35^\circ$ & 1.00 \\
1.28 & 250~$\mu$m & 771 & $33^\circ$ & 0.74 \\
1.33 & 280~$\mu$m & 786 & $35^\circ$ & 0.50 \\
1.51 & 285~$\mu$m & 764 & $33^\circ$ & 0.26 \\
3.06 & 280~$\mu$m & 923 & $34^\circ$ & 0.00 \\
\end{tabular}
\end{center}
\caption{
Details of the six emulsion experiments that measured clogging
probabilities.  $w$ is the hopper exit width, $d$ the mean droplet
diameter, $N$ is the number
of droplets, $\theta$ is the hopper angle, and $P_{\rm clog}$ is the
probability of clogging based on 50 trials.  The uncertainty of
$d$ is $\pm 5$~$\mu$m, and the uncertainty of $w/d$ is $\pm 0.03$.
}
\label{pj}
\end{table}

Note a caveat:  the more droplets that flow through, the more chance
there is to observe clogging, if the probability of clogging per
droplet is nonzero \cite{to05,janda08,tang09,lafond13}.  We cannot
perfectly control the number of droplets in our sample chamber,
so the cases with more droplets will have $P_{\rm clog}$ larger.
For the three points with $0 < P_{\rm clog} < 1$, the number of
droplets is fairly similar (see Table~\ref{pj}).  In the first
experiment reported by To {\it et al.} they used 200 particles
\cite{to01}, approximately a quarter of the number we use.
Their later work showed that with more particles $P_{\rm clog}$
moves to larger $w/d$ \cite{to02}.  They found $P_{\rm clog} =
1/2$ at $w/d \approx 4.0$ for 200 particles, and $\approx 4.8$
for 700 particles.  Janda {\it et al.} found qualitatively similar
results in their 2D granular experiment, with $P_{\rm clog} = 1/2$
increasing from $w/d \approx 3$ with 50 particles to $w/d \approx
5.5$ with 50000 particles \cite{janda08}.

We fit our data to a sigmoidal function as shown in
Fig.~\ref{probplot}(a).  This finds a width $\approx 0.2$,
slightly smaller than the widths $\approx 0.3$ in Ref.~\cite{to01}.
It is not clear that the sigmoidal fit we use is correct; To {\it
et al.}~used a different fit, and their data with gear-shaped
particles had a decidedly non-sigmoidal shoulder \cite{to01}.
Likewise Janda {\it et al.} used a different fit \cite{janda08}.
Our data are not sufficient to distinguish subtle differences
in these fits, so we stick with the simple sigmoidal fit.

Figure \ref{pictures}(a,b) shows two examples of clogged samples.
Panel (a) shows $w/d \approx 0.8$ and a situation where the
influence of surface tension is weak enough that one droplet can
deform and slip through.  However, after that first droplet, the
remainder clog.  Panel (b) shows a small ``arch'' of two particles
that clog at $w/d \approx 1.0$.

\subsection{Hydrogel experiments}

The hydrogel experiments are done in a similar fashion to the
emulsion experiments.  We load the hopper with 200 particles and
then allow them to flow through the hopper.  We repeat this 20
times and compute $P_{\rm clog}$ from the fraction of times that
we observe clogging.  We do this for a variety of hopper opening
widths $w$ and also tilt angles $\theta$ ($\theta = 10^\circ,
20^\circ, 43^\circ, 90^\circ$).  The results are shown in
Fig.~\ref{probplot}(b).  As with the emulsions, increasing the
hopper opening width decreases $P_{\rm clog}$ for a fixed
gravitational force.
However, clogging is easier than for the emulsion
droplets.  We observe most clogs are due to arches with three
particles; at the lowest hopper openings, occasionally arches are
formed with only two particles, more similar to the emulsion
case.  For the largest hopper openings ($w/d \approx 2.4$),
occasionally arches form with four particles; an example is shown
in Fig.~\ref{pictures}(c).

The benefit of the hydrogel experiments is that the influence of
gravity is apparent:  clogging is easier for reduced gravity, as
signified by the curves shifting to the right.  The location
where $P_{\rm clog} = 1/2$ changes from $w/d=1.76$ to 2.37 as
gravity decreases by a factor of 6.  For smaller gravitational
forces, particles are moving slower when they first encounter the
hopper walls, although in all experiments particles quickly slow
down as they fill up the hopper and begin draining through the
exit opening.  When an arch is formed and the hopper clogs, we
notice that the particles in the arch are clearly more deformed
when gravity is large and/or when more particles remain in the
hopper trapped above the arch.  The deformation, along with the
increasing $P_{\rm clog}$ with decreasing gravity, suggests that
particle softness plays an important role in the clogging process.

\subsection{Simulation}
\label{simresults}

We find that in our two experimental systems of soft nearly
frictionless particles, the probability of clogging in hopper
flow is greatly reduced from prior published experiments that
studied hard frictional particles \cite{to01,to02,janda08}.
In our experiment, we only see clogging with exit apertures
significantly smaller than previously seen with frictional particles
\cite{deming29,brown58,beverloo61,nedderman82,sheldon10,wilson14}.

The simulations
allow us to vary the relative importance of gravity and contact
forces over a larger range than the experiments.  This is done
through the ratio $g/F_0$ (which is nondimensional; see
Sec.~\ref{simulations}).  As with the experiments, for
each simulation we initialize the droplets in random positions
above the hopper, let them fall, and observe if they completely
flow out of the hopper or if they clog.  We do $n=100$ runs for each
condition to measure $P_{\rm clog}$, the fraction of runs that clog.

For a moderate value $g/F_0 = 10^{-2}$, the simulation clogging
probability curve looks qualitatively similar to the experiments
[circles in Fig.~\ref{probplot}(c)].  Varying $g/F_0$ significantly
shifts the clogging probability curve in Fig.~\ref{probplot}(c),
from $g/F_0=10^{-1}$ (diamonds) to $g/F_0=10^{-4}$ (squares).
This confirms the significant role deformability plays in the
clogging process, here for data where friction is not present.

Figure \ref{pictures}(d,e) shows examples of arches found in
the simulations.  For the largest value of gravity, clogging is
most typically due to one large droplet that reaches the exit when
most other droplets have already exited [Fig.~\ref{pictures}(d)].
This is analogous to a droplet such as the large one shown in
Fig.~\ref{bigsqueeze}, but with fewer droplets above it such that
the driving pressure is not large enough to cause the large droplet
to deform.  Thus, the clogging probability curve for such a large
value of $g/F_0$ [green diamonds in Fig.~\ref{probplot}(c)] has
little to do with arch formation and more to do with the likelihood
of an unusually large droplet being one of the last ones left in
the hopper.  Figure
\ref{pictures}(e) shows the more interesting case for
a lower value of $g/F_0$ corresponding to weaker gravity (or
equivalently, stiffer droplets).  Large arches can form (up to
5 droplets) without requiring friction to be present.  This is
perhaps an unsurprising result, as the theory of To {\it et al.}
that explains their data does not require friction \cite{to01}.

\section{Discussion}
\label{compare}

\begin{figure}
\includegraphics[width=8cm]{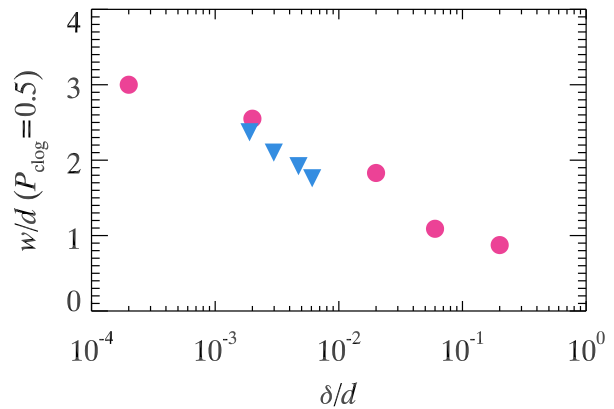}
\caption{
A plot of the size of the hopper opening $w/d$ for which $P_{\rm
clog} = 1/2$ as a function of $\delta/d$, the fractional
deformation of a particle due to its own weight.  (This is equal
to $2g/F_0$ for the simulation data; see text for discussion.)
The circles are simulation
data and the triangles are from the hydrogel data.
The left side of the graph corresponds to lower gravity or
stiffer particles.  A prior experiment with 200 steel disks
found $w/d=3.7$ \cite{to01}.
}
\label{halfprob}
\end{figure}

We can compare the experimental hydrogel
data and the simulation data.  
From the data shown in Fig.~\ref{probplot}(b,c), we
extract the hopper opening width $w/d$ for which $P_{\rm clog} =
0.5$.  To match experiment and simulation we consider the
magnitude of deformation $\delta/d$ a particle has due to its own
weight (nondimensionalized by particle diameter $d$).  
We compute this for hydrogel particles by 
balancing the weight of one particle with the Hertz contact
force law:
\begin{equation}
\frac{1}{6} \pi d^3 \rho g = \frac{4}{3} E^* \left( \frac{d}{2}
\right)^{1/2} \delta^{3/2}
\end{equation}
using the particle diameter $d = 13.1$~mm, density $\rho \approx
1$~g/cm$^3$, $g=9.8$~m/s$^2$, and $E^* = E / (1-\nu^2)$ in terms
of the Young's modulus and Poisson ratio (Sec.~\ref{hydrogel}).
Solving this we find that $\delta / d$ ranges from
0.006 to 0.002 as gravity goes from maximal to minimal (when we
tilt the chamber).  In other words, a single hydrogel particle only
deforms minimally due to gravity.  A similar calculation applied
to the bubble model through a balance of Eqns.~\ref{walleqn} and
\ref{graveqn} shows that in the bubble model $\delta / d = 2 g /
F_0$ (in the limit of small deformations).  These calculations
allows us to use $\delta/d$ to compare the simulation and
experimental data.  Figure~\ref{halfprob} shows the data for
the hydrogel experiments (triangles) and simulations (circles).
These results are in excellent agreement given that
there are no free parameters in the comparison.  In fact, given the
differences between the simulation (perfectly 2D, frictionless,
viscous interactions) and the hydrogel experiment, the agreement
is strong evidence that $\delta/d$ is a useful measure of the
importance of softness.  One neglected factor is that the
simulations used $N=800$ particles while the hydrogel experiments
used $N=200$; more particles in the hydrogel experiments likely
would increase the probability of clogging \cite{janda08} and thus
slightly raise the hydrogel data in Fig.~\ref{halfprob}.  
For a situation with a driving force other than gravity, a similar
parameter could be developed.  Note that $\delta/d$ is the
deformation of a particle due to its own weight; in a clogging
arch with particles supported by the arch, the deformation will
be significantly more.

We can compare the results of Fig.~\ref{halfprob} to the
experiments of To {\it et al.} that used steel disks \cite{to01}.
For gear-shaped particles, they found a slightly larger value,
$w/d=4.0$ for $P_{\rm clog}=0.5$ as compared to $w/d=3.7$ for the
smooth disks (using 200 disks).  Our results in Fig.~\ref{halfprob}
suggest that the relative influences of gravity and particle
stiffness (for example as quantified by the ratio $g/F_0$ in the
simulation) plays a more significant role for soft particles than
the enhanced friction played in the prior experiments.


Our results can also be compared with centrifuge experiments
of Dorbolo {\it et al.}, which found that clogging was
uninfluenced by gravity \cite{dorbolo13}.  
The difference between these experiments
and our work is likely explainable by their use of glass and
steel beads.  We can estimate the effective $\delta/d$ for
their experiment.  For glass beads, estimating their modulus as
$E=70$~GPa, density as $\rho=2.6$~g/cm$^3$, using their diameter
$d = 400$~$\mu$m, and their maximum imposed gravity ($20g$),
one finds at most $\delta/d \approx 10^{-6}$ for Ref.~\cite{dorbolo13}.
This is two decades lower $\delta/d$ than we have probed with our
simulations.  It is reasonable to conjecture that the experiments
of Ref.~\cite{dorbolo13} are still in a high particle stiffness
limit where the clogging results are independent of gravity --
equivalent to the low gravity limit of our softer particles, despite
the enhanced gravity of their experiments.  Comparing the results
of Fig.~\ref{halfprob} to the 
centrifuge experiments \cite{dorbolo13} suggests that $\delta /
d \lesssim 10^{-5}$ may be sufficient to reach a limit of hard
particles.

Our results are in qualitative agreement with soft frictional particle
simulations of Ar{\'e}valo and Zuriguel \cite{arevalo14,arevalo16},
who found that increasing gravity by four orders of magnitude
decreased the clogging probability slightly.  Their simulations
used stiffer particles that we considered, with $\delta/d =
10^{-8} - 10^{-4}$.  They simulated the case where the hopper was
kept continually full of particles and they measured the size of
avalanches in between clogging events, so a direct comparison with
Fig.~\ref{halfprob} is not possible.  As a rough comparison, for
$\delta/d=10^{-4}$ they found a mean avalanche size of $O(10^3)$
for an exit width $w/d = 4.0$ \cite{arevalo16}.  This is comparable
to the number of droplets in our simulation (800) and so $w/d=4.0$
seems roughly in agreement with the low $\delta/d$ limit of
Fig.~\ref{halfprob} as well as the steel particle results of To
{\it et al.} \cite{to01}.  Their data suggest that below $\delta /
d = 10^{-5}$ one should see little dependence of $P_{\rm clog}$
on $\delta / d$ \cite{arevalo16}.

For large $\delta / d$ we find clogging is difficult to observe in
our emulsion experiments [Fig.~\ref{probplot}(a)].  There we find
$P_{\rm clog} = 1/2$ at $w/d = 1.37$.  From Fig.~\ref{halfprob}
this corresponds to $\delta/d \approx 3 \times 10^{-2}$.  This
fractional deformation is consistent with visual observation of
isolated droplets.

\section{Conclusions}

In our experiments, our soft particles cannot sustain long arches, and
clogging requires small openings.  One possible explanation is the
lack of static friction in our experiment; our result of reduced
clogging qualitatively matches the trend seen by To {\it et al.}
who found a lower clogging probability for smooth-surface disks
compared to gear-shaped disks \cite{to01}.  However, our simulation
results show that even frictionless droplets can form large arches
under certain conditions [Fig.~\ref{pictures}(e)].  The key requirement
is that the gravitational force must be small in comparison to the
stiffness of the droplets.  To rephrase this in physical terms,
for maximal clogging
an emulsion droplet would need a high surface tension, a particle
needs a large elastic modulus, or the driving force (e.g.
gravity) must be low.  For
example in our emulsion experiment, despite the reduced influence of
gravity (due to buoyancy of the droplets) and their slower motion
(due to viscous forces), gravity essentially breaks large arches
due to a mechanism similar to what is seen in Fig.~\ref{bigsqueeze},
albeit with subtler droplet deformations.


Our clogging results with reduced gravity are the opposite of those seen
in prior work that found reducing forcing prevented clogging
\cite{helbing00,zuriguel14,pastor15}, and the reasons for this difference are
important.  The prior observation is termed ``faster-is-slower''
and was observed in simulations of pedestrians, where panic is
counterproductive to exiting a room through a small door
\cite{helbing00}.  In later work that studied clogging in a
variety of situations, the conclusion was that reducing
the load on the arches at the exit allows vibrations or other
noise to destroy the arch and thus the hopper flow can resume
\cite{zuriguel14,cates98}.  Or increasing the load,
the weight of the grains above the exit applies a compatible
load thus strengthening the arch and increasing the
persistence of the clog.
In contrast, our soft particles are deformed by this load
which can strengthen the arch (at low loads) or break the
arch (at high loads).  More significantly, we have no source of
incompatible forces that disrupt a stable arch once formed, unlike
the prior work \cite{zuriguel14}.  One could imagine a reversal
of our soft particle results by adding vibrations to our macroscopic
hydrogel experiment, or shrinking the oil droplet experiment so
that Brownian motion becomes significant; both mechanical and
thermal vibrations were shown to decrease clogging in the prior
work \cite{zuriguel14}.  To be clear, in both the prior work and
in our work, increasing the driving force increases the
outflow flux rate as long as the system is not clogged
\cite{arevalo14,arevalo16,ashour17}; the key difference is the
system behavior after a clogging arch is formed.

Overall, our results demonstrate that the flow of soft particles
is qualitatively different from the case of hard particles.
Hard particle behavior appears as a limiting case where the driving
is low or the particle stiffness is high, such that particles
are barely deformable during the flow.  Our results potentially
have implications for other situations where particles have soft
long-range interactions such as magnetic particles \cite{lumay15},
merging traffic \cite{helbing01}, and perhaps flowing bacteria
\cite{altshuler13}.  This may also explain why experiments with
ants found no clogging with higher driving force \cite{boari13}.

We thank D.~Chen, K.~Desmond, D.~Durian, C.~Roth, M.~Thees,
and I.~Zuriguel for helpful discussions, and J.~Burton,
N.~Cuccia, Y.~Gagnon, and C.~Orellana for help with the hydrogel
mechanical measurements.  This work was supported by the National
Science Foundation (X.H. supported by CBET-1336401, M.M. and
E.R.W. supported by DMR-1609763).

\end{document}